\documentclass{aa}
\usepackage{txfonts}
\usepackage{graphicx}
\usepackage{natbib}
\bibpunct{(}{)}{;}{a}{}{,}

\begin{document}

\title{Ages and metallicities of star clusters: new calibrations and
diagnostic diagrams from visible integrated spectra} 
\titlerunning{Ages and metallicities from star cluster spectra}

\author{J. F. C. Santos Jr.\inst{1} \and A. E. Piatti\inst{2}}

\institute{Departamento de F\'{\i}sica, ICEx, UFMG, CP 702, 30123-970 Belo 
Horizonte, MG, Brazil \and Instituto de Astronom\'{\i}a y Fisica del 
Espacio, CC 67, Suc. 28, 1428, Buenos Aires, Argentina}

\offprints{J. F. C. Santos Jr., \email{jsantos@fisica.ufmg.br}}

\date{Received / Accepted }

\abstract{We present homogeneous scales of ages and metallicities for star
clusters from very young objects, through intermediate-age 
ones up to the oldest known clusters. All the selected clusters 
have integrated
spectra in the visible range, as well as reliable determinations of their
ages and metallicities. From these spectra equivalent widths (EWs) 
of K\,Ca\,II, 
G\,band\,(CH) and Mg\,I metallic, and H$\delta$, H$\gamma$ and H$\beta$ 
Balmer lines have been measured homogeneously. The analysis of these EWs 
shows
that the EW sums of the metallic and Balmer H lines, separately, are good
indicators of cluster age for objects younger than 10 Gyr, and that the
former is also sensitive to cluster metallicity for ages greater than 10 Gyr.
We propose an iterative procedure for estimating cluster ages by employing
two new diagnostic diagrams and age calibrations based on the above EW sums.
For clusters older than 10 Gyr, we also provide a calibration to
derive their overall metal contents.

}

\maketitle

\section{Introduction}

The determination of age and metallicity of open and globular clusters have 
contributed to the present knowledge of the structure and chemical evolution of
our Galaxy. Properties of star cluster systems in external galaxies, including 
both Magellanic Clouds (MCs), are also good tracers of the 
galaxy's chemical enrichment history.

Age determinations of star clusters are frequently based on isochrone matching 
to color-magnitude diagrams whenever individual star photometry is 
possible, this technique thus being constrained to 
spatially resolved 
objects. Direct metallicity determinations from spectroscopic observations of
individual cluster stars also suffer from spatial resolution limitations 
and they are generally done for stellar systems in the Local Group.
Other approaches to estimate metallicity of clusters rely on the 
position and shape of evolved red features in their color-magnitude 
diagrams \citep[e. g.][]{gs99}. 

Integrated colors and spectra have also been used to rank clusters 
according to their age and metallicity. In particular, the Gunn system 
photometric classification by \cite{swb80} of star clusters in the 
MCs was later extended to Johnson $UBV$ photometric system in which 
age calibrations were obtained \citep{ef85,gcb95,bcd96}. \cite{r82} analyzed a 
sample of integrated spectra of red star clusters in the MCs proposing a 
diagnostic diagram involving the equivalent widths (EWs) of Balmer lines and 
that of K\,Ca\,II. This diagram discriminated most of the MCs 
clusters from the Galactic globulars, indicating age (and metallicity to a 
lesser degree) as the main segregation parameter. The lack of a large 
sample of clusters with well-determined ages and metallicities prevented 
Rabin from exploring an empirical calibration of the data, although the diagram provided a means to estimate approximate cluster ages from measured EWs in the 
integrated spectra. Aiming at population synthesis studies, \cite{ba86a} 
gathered a library of star cluster spectra in which age and metallicity trends
were searched for by plotting EWs of various spectral features against
age and metallicity. They show that the EW of each Balmer line is a 
bivalued function of age with a maximum around 300\,Myr and that the EWs
of prominent metallic features are metallicity indicators for old clusters.
Their original sample, composed of 19 MC clusters, 41 Galactic globulars and 3 
open clusters, has been substantially expanded over the years in both spectral
range and number of objects
\citep[][ and references therein]{sab02,pbc02}, and 
is available by internet 
(http://vizier.u-strasbg.fr/cgi-bin/VizieR?-source=III/219).

Recently, a large number of extragalactic star cluster systems has been
investigated thanks to the increasing sensitivity of observational technology
\citep[see][ and references therein]{wcm04}.
Even so, only integrated light can be gathered for the more distant systems. 
By means of such studies it has been possible to examine the origin and 
evolution of the parent galaxy, as well as to learn about our own Galaxy 
in a comparative way. On the other hand, the calibration of fundamental 
properties of our local cluster systems against measured indices on integrated 
light observations is a crucial step to get information on extragalactic 
cluster systems.

The present work is intended to provide a useful tool for diagnosing the 
age and the metallicity of single burst stellar populations. The method 
rests on the measurement of EWs in the integrated spectra of star clusters 
and on the parameters (age and metallicity) taken from the literature put 
in homogeneous scales. The large dataset for  which age and 
metallicity estimates are available make this a worthwhile analysis. This 
paper is structured as follows. In Sect. 2 the database is presented, 
while in Sect. 3 the EW measurements for selected features are detailed. The 
calibration of the EWs as a function of age and metallicity is described in 
Sect. 4 and the diagnostic diagrams are discussed in Sect. 5. Concluding 
remarks are given in Sect. 6.  

\section{Homogeneous scales of age and metallicity}

The sample selection was based on the availability of cluster integrated 
spectra with estimated ages and metallicities. Aiming at homogeneity of the 
sample properties, ages and metallicities were transformed towards a 
uniform scale, starting with the Galactic globular clusters (GGCs). These 
procedures have already been used to discuss a possible reorganization of the 
original template spectra of \cite{b88}, by grouping GGCs in different 
bins, using as criteria, in addition to the metallicity, the age and the 
horizontal branch morphology \citep{sab02}.

\subsection{The Galactic globular cluster metallicity scale}

The metallicity scale defined by \citet[][ hereafter CG97]{cg97} was 
preferred over the widely used \citet[][ hereafter ZW84]{zw84} scale, 
because the former is based on high-dispersion CCD spectra of giants in 24 
clusters, characterizing an homogeneous sample. The metallicities presented by 
\citet[][ hereafter R97]{rhs97}, which were obtained from the 
near-infrared Ca\,II triplet ($\lambda$8498, 8542, 8662\AA), were adopted in 
the present study. Twenty-three globular clusters in our sample are not in 
R97's sample and their metallicities were taken from ZW84 and transformed to 
CG97's metallicity scale. The exceptions were metal-rich clusters (see below) 
and \object{NGC\,6540}, not present in either sample, for which the quoted metallicity 
in the \cite{h96} compilation was adopted (from the 2003 updated 
version of the catalogue available at 
http://physwww.physics.mcmaster.ca/\~\,harris/mwgc.dat).

Concerning clusters for which there are metallicities available in the 
ZW84 scale, their [Fe/H] were transformed to the CG97 scale by using eq.~7 in 
CG97:

\begin{equation}
\label{fe}
{\rm [Fe/H]}_{CG}=a+b{\rm [Fe/H]}_{ZW}+c{\rm [Fe/H]}_{ZW}^2  
\end{equation}

\noindent where $a=-0.618\pm0.083$, $b=-0.097\pm0.189$, $c=-0.352\pm0.067$ 
and $\sigma$(${{\rm [Fe/H]}_{CG}}$)$=0.08$ for 24 clusters. The relationship is
valid in the range $-2.24<{\rm [Fe/H]}_{ZW}<-0.51$ and therefore it was not 
applied to the metal-rich globular clusters. The metallicity errors were 
propagated from their original ZW84 values for individual clusters.

\subsubsection{The metal-rich GGCs}

There are discrepancies (larger than $\approx0.3$\,dex) between the 
aforementioned metallicity scales for the metal-rich clusters 
([Fe/H]$_{ZW}>-0.5$). Specifically, the metallicities of \object{NGC\,6316}, 
\object{NGC\,6440}, 
\object{NGC\,6528}, \object{NGC\,6553}, \object{NGC\,6624} and 
\object{NGC\,6637} were reanalyzed in terms of more 
recent studies in the literature, a task whose fundamental issue was to keep 
the final adopted metallicities as consistent as possible. The [Fe/H] 
discrepancies are probably produced by uncertainties in the EWs, since 
high line crowding in red giant spectra can affect the continuum placement. 

\cite{cgb99} obtained high-resolution near-infrared spectra of 5 horizontal 
branch (HB) stars in \object{NGC\,6553}, and estimated a metallicity of 
[Fe/H]$_{CG}=-0.16\pm0.08$ from uncrowded individual spectral lines, which is 
much higher than the value quoted by R97 ([Fe/H]$_{CG}=-0.60\pm0.04$). Later, 
\cite{car01} revised Cohen et al.'s estimate to [Fe/H]$_{CG}=-0.06\pm0.15$. We 
adopted this last estimate. Having as one of their goals to extend the 
calibration of the CG97 metallicity scale to the metal-rich regime, 
\cite{car01}
obtained [Fe/H]$_{CG}=0.07\pm0.10$ for \object{NGC\,6528} from high-resolution spectra 
of 4 HB member stars. They propose a new calibration to transform the ZW84 
scale to the CG97 one in order to account for the metal-rich clusters. We have 
not used such a new calibration because there still seems to remain 
important discrepancies in the metal-rich regime. For instance, if one 
assumes [Fe/H]$_{ZW}=0.12\pm0.21$ for \object{NGC\,6528} and uses this new 
calibration \citep[eq.~3 in][]{car01}, it gives
[Fe/H]$_{CG}=1.1\pm0.7$. Instead, metallicities for \object{NGC\,6528} and 
\object{NGC\,6553} were adopted from the direct measurements on high-dispersion 
spectra according to the analyses in the literature mentioned above. 

\object{NGC\,6316} ([Fe/H]$_{ZW}=-0.47\pm0.15$) and \object{NGC\,6440} 
([Fe/H]$_{ZW}=-0.26\pm0.15$) have their  metallicities revised 
to lower values 
([Fe/H]$_{ZW}=-0.55\pm0.11$ and [Fe/H]$_{ZW}=-0.34\pm0.11$, respectively, 
according to the \cite{az88} analysis of integrated near-infrared spectra). 
Subsequently, \cite{m95} determined [Fe/H]$_{ZW}=-0.50\pm0.20$ for 
\object{NGC\,6440} 
based on the spectra of 11 cluster giants together with their near-infrared 
colors. The adopted metallicity for this cluster was obtained by entering this 
value into eq.~\ref{fe} giving [Fe/H]$_{CG}=-0.66\pm0.14$, from a slight 
extrapolation. In the case of \object{NGC\,6316}, the
mean metallicity from ZW84 and \cite{az88} values ([Fe/H]$_{ZW}=-0.51\pm0.19$) 
transformed to CG97 scale by using eq.~\ref{fe} 
([Fe/H]$_{CG}=-0.66\pm0.14$) was adopted in the following analysis. 

\cite{hjz00} have obtained spectra in the Ca\,II triplet region of 4 members of
\object{NGC\,6624} and 7 members of \object{NGC\,6637}, and 
estimated their [Fe/H]$_{ZW}$. These 
values were transformed to  [Fe/H]$_{CG}$ according to  eq.~\ref{fe} and 
compared to the ones given by R97. Within the errors, the metallicities are 
similar: for \object{NGC\,6624}, [Fe/H]$_{CG}=-0.70\pm0.09$ according 
to \cite{hjz00}, 
and $-0.70\pm0.03$ following R97; for \object{NGC\,6637} the values 
are $-0.70\pm0.09$ 
and $-0.78\pm0.03$, respectively.

\subsection{The Galactic globular cluster age scale}

The Galactic globular cluster relative age calibration by 
\citet[ hereafter R99]{rsp99}, based on uniform $VI$ CCD color-magnitude 
diagrams and different sets of isochrones, was adopted. Absolute ages were 
established with the  post-Hipparcos calibration by \cite{cgc00}, yielding an 
average absolute age of 13.2\,Gyr. An uncertainty of 0.1\,dex in [Fe/H] 
corresponds to an age precision within 0.25\,Gyr (R99). Since just 9 out of 42 
globular clusters in our sample have ages in the R99 scale, an age-metallicity 
relation has been used in order to get ages  for 25 
clusters with [Fe/H] $\le$ -0.7 dex. Specifically, a 2nd order polinomial
was fitted to age as a function 
of metallicity in the CG97 scale for the 35 clusters in R99 sample, resulting
in:

{\small
\begin{equation}
\label{tfe}
{\rm t(Gyr)}=4.3(\pm2.2)-11.0(\pm3.3){\rm [Fe/H]}_{CG}-3.4(\pm1.1){\rm [Fe/H]}_{CG}^2
\end{equation}
}

For a given cluster [Fe/H]$_{CG}$, the cluster age and its error were 
estimated from eq.~\ref{tfe} and its dispersion at that metallicity, 
respectively. For the metal-rich clusters with [Fe/H]$_{CG}>-0.7$ we assigned 
ages of $10.0\pm2.0$\,Gyr, which correspond to an extrapolated value.

Table~\ref{tab1mw} presents the final adopted parameters for the Galactic
globular clusters with observed integrated spectra.

\begin{table}
\tiny
\caption[]{Homogeneous ages and metallicities: Milky Way clusters.}
\label{tab1mw}
\begin{tabular}{lrrrrrrrrrrr}
\hline
CLUSTER&[Fe/H]$_{\rm CG}$&$\sigma$([Fe/H])&Ref.&t(Gyr)&$\sigma$(t)&Ref.\\
\hline
GLOBULAR\\
\hline
\object{NGC\,104}     & -0.78  & 0.02 & 1 & 11.9  & 1.0   &7\\
\object{NGC\,362}     & -1.09  & 0.03 & 1 & 10.2  & 1.0   &7\\
\object{NGC\,1851}     & -1.03  & 0.06 & 1 & 10.6  & 0.9   &7\\
\object{NGC\,1904}     & -1.37  & 0.05 & 1 & 13.2  & 1.1   &7\\
\object{NGC\,2808}     & -1.11  & 0.03 & 1 & 10.7  & 0.9   &7\\
\object{NGC\,4590}     & -2.00  & 0.03 & 1 & 12.4  & 1.1   &7\\
\object{NGC\,4833}     & -1.71  & 0.03 & 1 & 13.4  & 1.0   &8\\
\object{NGC\,5024}     & -1.88  & 0.49 & 2 & 13.2  & 1.0   &8\\
\object{NGC\,5824}     & -1.67  & 0.47 & 2 & 13.4  & 1.0   &8\\
\object{NGC\,5927}     & -0.64  & 0.02 & 1 & 10.0  & 2.0   &9\\
\object{NGC\,5946}     & -1.15  & 0.33 & 2 & 12.5  & 2.0   &8\\
\object{NGC\,6093}     & -1.47  & 0.04 & 1 & 13.7  & 0.9   &7\\
\object{NGC\,6139}     & -1.42  & 0.40 & 2 & 13.2  & 1.0   &8\\
\object{NGC\,6171}     & -0.95  & 0.04 & 1 & 13.5  & 0.9   &7\\
\object{NGC\,6287}     & -1.90  & 0.53 & 2 & 13.2  & 1.0   &8\\
\object{NGC\,6293}     & -1.73  & 0.48 & 2 & 13.4  & 1.0   &8\\
\object{NGC\,6304}     & -0.66  & 0.03 & 1 & 10.0  & 2.0   &9\\
\object{NGC\,6316}     & -0.66  & 0.14 & 2,3 & 10.0  & 2.0   &9\\
\object{NGC\,6356}     & -0.69  & 0.16 & 2 & 10.0  & 2.0   &9\\
\object{NGC\,6388}     & -0.74  & 0.18 & 2 & 10.6  & 2.0   &8\\
\object{NGC\,6401}     & -0.96  & 0.27 & 2 & 11.8  & 2.0   &8\\
\object{NGC\,6402}     & -1.16  & 0.32 & 2 & 12.6  & 2.0   &8\\
\object{NGC\,6440}     & -0.66  & 0.14 & 4 & 10.0  & 2.0   &9\\
\object{NGC\,6453}     & -1.29  & 0.37 & 2 & 12.9  & 1.0   &8\\
\object{NGC\,6517}     & -1.12  & 0.32 & 2 & 12.5  & 2.0   &8\\
\object{NGC\,6528}     &  0.07  & 0.10 & 5 & 10.0  & 2.0   &9\\
\object{NGC\,6540}     & -1.2   & 0.5  & 6 & 12.7  & 2.0   &8\\
\object{NGC\,6541}     & -1.53  & 0.03 & 1 & 13.3  & 1.0   &8\\
\object{NGC\,6544}     & -1.20  & 0.04 & 2 & 12.7  & 1.5   &8\\
\object{NGC\,6553}     & -0.06  & 0.15 & 5 & 10.0  & 2.0   &9\\
\object{NGC\,6558}     & -1.21  & 0.34 & 2 & 12.7  & 1.5   &8\\
\object{NGC\,6569}     & -0.79  & 0.20 & 2 & 10.9  & 2.0   &8\\
\object{NGC\,6624}     & -0.70  & 0.03 & 1 & 10.4  & 1.5   &8\\
\object{NGC\,6637}     & -0.78  & 0.03 & 1 & 10.9  & 2.0   &8\\
\object{NGC\,6638}     & -0.90  & 0.04 & 1 & 11.5  & 2.0   &8\\
\object{NGC\,6642}     & -1.08  & 0.31 & 2 & 12.3  & 2.0   &8\\
\object{NGC\,6652}     & -0.81  & 0.21 & 2 & 11.1  & 2.0   &8\\
\object{NGC\,6715}     & -1.25  & 0.07 & 1 & 12.9  & 1.5   &8\\
\object{NGC\,6760}     & -0.66  & 0.14 & 2 & 10.0  & 2.0   &9\\
\object{NGC\,6864}     & -1.10  & 0.30 & 2 & 12.4  & 2.0   &8\\
\object{NGC\,7006}     & -1.35  & 0.36 & 2 & 13.1  & 1.0   &8\\
\object{NGC\,7078}     & -2.02  & 0.04 & 1 & 12.9  & 0.6   &7\\
\hline
\end{tabular}
\end{table}
\setcounter{table}{0}

\begin{table}
\tiny
\caption[]{Homogeneous ages and metallicities: Milky Way clusters (cont.)}
\begin{tabular}{lrrrrrrrrrrr}
\hline
CLUSTER&[Fe/H]$_{\rm CG}$&$\sigma$([Fe/H])&Ref.&t(Gyr)&$\sigma$(t)&Ref.\\
\hline
OPEN\\
\hline
\object{NGC\,2158    } & -0.25  & 0.09 & 10 &  2.0  & 0.5   &17\\
\object{vdB-RN\,80   } &  0.0   & 0.2  & 11 & 0.0045& 0.0015&16\\
\object{NGC\,2368    } &  0.0   & 0.2  & 11 &  0.05 & 0.01  &16\\
\object{Berkeley\,75 } &  0.0   & 0.2  & 11 &  3.0  & 1.0   &16\\
\object{Haffner\,7   } &  0.0   & 0.2  & 11 &  0.10 & 0.01  &16\\
\object{ESO\,429-SC13} &  0.0   & 0.2  & 11 &  0.10 & 0.05  &16\\
\object{NGC\,2660    } & -0.18  & 0.06 & 12 &  1.1  & 0.1   &16\\
\object{UKS\,2       } &  0.0   & 0.2  & 11 &  0.8  & 0.2   &16\\
\object{Ruprecht\,83 } &  0.0   & 0.2  & 11 &  0.055& 0.020 &16\\
\object{Hogg\,3      } &  0.0   & 0.2  & 11 &  0.075& 0.025 &16\\
\object{NGC\,3293    } &  0.0   & 0.2  & 11 &  0.006& 0.001 &16\\
\object{Bochum\,12   } &  0.0   & 0.2  & 11 &  0.045& 0.015 &16\\
\object{Pismis\,17   } &  0.0   & 0.2  & 11 & 0.0045& 0.0015&16\\
\object{Hogg\,11     } &  0.0   & 0.2  & 11 &  0.008& 0.005 &16\\
\object{ESO\,93-SC08 } & -0.4   & 0.2  & 13 &  5.5  & 1.0   &13\\
\object{MEL\,105     } &  0.00  & 0.25 & 14 &  0.3  & 0.05  &16\\
\object{BH\,132      } &  0.0   & 0.2  & 11 &  0.15 & 0.05  &16\\
\object{Hogg\,15     } &  0.0   & 0.2  & 11 &  0.02 & 0.01  &16\\
\object{Pismis\,18   } &  0.0   & 0.2  & 11 &  1.2  & 0.4   &16\\
\object{NGC\,5606    } &  0.09  & 0.25 & 14 &  0.006& 0.002 &16\\
\object{NGC\,5999    } &  0.0   & 0.2  & 11 &  0.3  & 0.1   &16\\
\object{NGC\,6031    } &  0.0   & 0.2  & 11 &  0.2  & 0.1   &16\\
\object{Ruprecht\,119} &  0.0   & 0.2  & 11 &  0.015& 0.010 &16\\
\object{NGC\,6178    } &  0.0   & 0.2  & 11 &  0.04 & 0.01  &16\\
\object{Lyng\aa\,11  } &  0.0   & 0.2  & 11 &  0.45 & 0.05  &16\\
\object{NGC\,6253    } &  0.5   & 0.1  & 15 &  3.0  & 0.5   &15\\
\object{BH\,217      } &  0.0   & 0.2  & 11 &  0.020& 0.015 &16\\
\object{NGC\,6318    } &  0.0   & 0.2  & 11 &  0.02 & 0.02  &16\\
\object{NGC\,6520    } & -0.25  & 0.25 & 14 &  0.19 & 0.04  &16\\
\object{NGC\,6603    } &  0.0   & 0.2  & 11 &  0.35 & 0.10  &16\\
\object{Ruprecht\,144} &  0.0   & 0.2  & 11 &  0.15 & 0.05  &16\\
\object{NGC\,6705    } &  0.14  & 0.04 & 12 &  0.25 & 0.05  &16\\
\object{NGC\,6756    } &  0.0   & 0.2  & 11 &  0.3  & 0.1   &16\\
\hline
\end{tabular}

References: (1) \cite{rhs97}, (2) \cite{zw84} plus 
eq.~\ref{fe}, (3) \cite{az88} plus eq.~\ref{fe}, (4) 
\cite{m95} plus eq.~\ref{fe}, (5) \cite{car01}, (6) \cite{h96},
(7) \cite{rsp99}, (8) eq.~\ref{tfe}, (9) assumed age for metal-rich 
GGCs (see Sect. 2.2), (10) \cite{fri02}, (11) assumed [Fe/H] 
for open clusters (see Sec 2.3), (12) \cite{taa97}, (13) \cite{ps03}, 
(14) \cite{t03}, (15) \cite{tal03}, (16) \cite{pbc02},
(17) \cite{cgm02}.
\end{table}

\begin{table}
\tiny
\caption[]{Homogeneous ages and metallicities: Magellanic Cloud clusters.}
\label{tab1mc}
\begin{tabular}{lrrrrrrrrrrr}
\hline
CLUSTER&[Fe/H]$_{\rm CG}$&$\sigma$([Fe/H])&Ref.&t(Gyr)&$\sigma$(t)&Ref.\\
\hline\hline
LMC\\
\hline
\object{NGC\,1466}     & -1.64  & 0.49 & 18,37 & 13.1  & 1.5   &29\\
\object{NGC\,1711}     & -0.68  & 0.15 & 19 &  0.068& 0.009 &30\\
\object{NGC\,1783}     & -0.65  & 0.14 & 20 &  1.3  & 0.4   &35\\
\object{NGC\,1805}     & -0.2   & 0.2  & 21 &  0.014& 0.006 &31\\
\object{NGC\,1831}     & -0.62  & 0.09 & 23 &  0.32 & 0.12  &30\\
\object{NGC\,1850}     & -0.12  & 0.03 & 22 &  0.031& 0.009 &32\\
\object{NGC\,1854}     & -0.50  & 0.10 & 22 &  0.034& 0.008 &32\\
\object{NGC\,1856}     & -0.17  & 0.27 & 25 &  0.151& 0.040 &32\\
\object{NGC\,1866}     & -0.66  & 0.14 & 24 &  0.15 & 0.05  &30\\
\object{NGC\,1868}     & -0.66  & 0.14 & 18 &  0.85 & 0.11  &30\\
\object{NGC\,1978}     & -0.85  & 0.24 & 24 &  2.2  & 0.4   &24\\
\object{NGC\,1984}     & -0.90  & 0.40 & 26 &  0.004& 0.004 &33\\
\object{NGC\,2004}     & -0.56  & 0.03 & 22 &  0.028& 0.018 &30\\
\object{NGC\,2011}     & -0.47  & 0.40 & 26 &  0.005& 0.001 &32\\
\object{NGC\,2100}     & -0.32  & 0.03 & 22 &  0.032& 0.019 &30\\
\hline
SMC\\
\hline
\object{NGC\,121 }     & -1.19  & 0.12 & 27 & 11.9  & 1.3   &36\\
\object{NGC\,330 }     & -0.82  & 0.10 & 27 &  0.025& 0.015 &27\\
\object{NGC\,419 }     & -0.70  & 0.30 & 27 &  1.2  & 0.5   &27\\
\object{K\,3     }     & -0.98  & 0.12 & 27 &  6.0  & 1.3   &36\\
\object{K\,28    }     & -1.2   & 0.2  & 28 &  2.1  & 0.5   &28\\
\hline
\end{tabular}

References: (18) \cite{oss91}  and eq.~\ref{fe},
(19) \cite{drg00} and eq.~\ref{fe}, (20) \cite{c82}  and 
eq.~\ref{fe}, (21) \cite{jbg01}, (22) \cite{jt94},
(23) \cite{lr03}, (24) \cite{hfs00} and
eq.~\ref{fe}, (25) \cite{bhs02}, (26) \cite{oo98}, 
(27) \cite{dh98}, (28) \cite{psc01}, (29) \cite{jbs99} (see Sect. 2.3),
(30) \cite{gcb95}, (31) \cite{dgj02}, (32) \cite{bas90}, (33) 
\cite{sbc95}, (34) \cite{psg02}, (35) 
\cite{gbd97}, (36) \cite{msf98}, (37) \cite{sso92} and eq.~\ref{fe}.
\end{table}

\subsection{The Galactic open and Magellanic Cloud clusters}

Metallicities and ages were assigned to Galactic open (GOCs) and 
Magellanic Cloud clusters (MCCs) with observed spectra available. 
In order to check how smooth the link is between the properties of the oldest 
clusters in different environments and those of intermediate-age/young 
clusters, the scales adopted for the younger GGCs were considered. Indeed, 
\object{NGC\,1466} and \object{NGC\,6253} allowed us to perform 
such a comparison.
Inevitably, a large number of works on the determination of GOC and MCC 
properties are based on different observational techniques and methods of 
analysis. Our attempt to homogenize the cluster properties gathered from the 
literature is in the hope that the relationship between the observational
quantities and the fundamental properties of star systems (e. g., age and 
metallicity) is not degraded by the lack of consistency among these 
properties. 

For a given cluster, the comparison between ages and/or metallicities
estimated from different methods guided us to achieve the final 
homogeneous dataset. The general guidelines applied for adopting the final
parameters were as follows: metallicities obtained from spectroscopic methods 
were preferred over photometric ones, and whenever the case, its mean 
value was brought to the same GCC scale according to eq.~\ref{fe}. Isochrone 
matching to CMDs and spectral flux distributions were the methods 
selected in the literature for deciding the final adopted ages. The
final adopted parameters of Galactic open and Magellanic Cloud clusters with 
observed spectra are given in Table~\ref{tab1mw} and Table~\ref{tab1mc}, 
respectively. Below, we describe the intercomparison between different 
studies for clusters deserving some comments, which illustrates the 
process of the parameter merging employed.

The metallicity of \object{NGC\,6253} derived by \cite{tal03} using Str\"omgren
photometric indices and $Ca$ and H$_\beta$ filters was adopted in the 
present work. Its age was also adopted from \cite{tal03}, who transformed 
$b-y$ colors to $B-V$ colors in order to obtain a suitable CMD for 
isochrone matching (Padova and Geneva models used). We note that \cite{at00} 
obtained [Fe/H] = $-1.82\pm0.04$ and age = $12.0\pm0.8$ Gyr for 
\object{NGC\,6397} using
the same kind of data and approach as \cite{tal03}. In our adopted scales, 
\object{NGC\,6397} has [Fe/H] = $-1.76\pm0.03$ and 
age = $13.2\pm0.9$ Gyr, which 
essentially are, within the uncertainties, the same values as 
obtained by \cite{at00}.

For \object{Melotte\,105}, \object{NGC\,5606} and \object{NGC\,6520}, 
the metallicities obtained by 
\cite{t03} from the $UV$ excess method \citep{c85} were 
adopted, since these clusters lack spectroscopy based estimates. The 
uncertainty from the original work by \cite{c85}, namely $\pm0.25$ in [Fe/H],
is such that it should encompass different scales.

 Concerning MCCs, \cite{oss91} have determined [Fe/H]$_{ZW}$ for 
4 clusters in our 
sample (\object{NGC\,1466}, \object{NGC\,1831}, \object{NGC\,1868}, 
\object{NGC\,1978}), from measurements of 
Ca\,II near-infrared triplet EWs.

\cite{gbd97} calibrated the magnitude difference $\delta$$T_1$ between
the giant branch clump and the turnoff in terms of age using LMC and
Galactic standard clusters older than 1 Gyr. The mean ages of 7 LMC standard 
clusters older than $\approx$ 10 Gyr is $<$age(Gyr)$> = 14.0\pm0.9$. 
\cite{ohm98} observed another 6 old LMC clusters and provided relative
ages 
by using the GGCs \object{M\,3}, \object{M\,5} and \object{M\,55} 
as age standards. We computed the mean 
age of these 6 LMC clusters from the absolute ages of \object{M\,3}, 
\object{M\,5} and \object{M\,55} 
according to R99 and \cite{cgc00} and obtained $<$age(Gyr)$> = 14.0\pm1.4$. 
Since the mean ages of the old objects in the \cite{gbd97} and \cite{ohm98} 
samples are in very good agreement, we deduce that, on average, the age scale 
by \cite{gbd97} is compatible with the scale adopted in the present study.

The following analysis of \object{NGC\,1466} reinforces such a compatibility.
The cluster [Fe/H]$_{ZW}$, as derived by \cite{oss91} and revised by 
\cite{sso92}, aiming at consistency with RR Lyrae analyses, was transformed to 
[Fe/H]$_{CG}$. \cite{jbs99} used this metallicity value and the cluster 
HST CMD to derive its age. On the basis of isochrone matching, they obtained 
the same age value as that of \object{M\,3} and \object{M\,92}, with 
an error of 
1.5\,Gyr. Since R99 provide  identical ages for these GGCs 
(13.1\,Gyr), we use it for \object{NGC\,1466}. This age incorporates the value 
12.7\,Gyr determined by \cite{gbd97} from the $\delta$$T_1$ age index. 

\cite{gbd97} also measured ages for \object{NGC\,1783} and \object{NGC\,1978}. 
We adopted their age for \object{NGC\,1783}, with errors that encompass 
previous works. In the case of \object{NGC\,1978}, instead, we adopted 
the age estimated by \cite{tfc99} from the
fit of different sets of isochrones with [Fe/H]$_{ZW}$ = -0.4. The metal 
content 
assumed by \cite{tfc99} is compatible with the more recent and reliable 
determination obtained by \cite{hfs00}. They employed high 
resolution spectra 
obtained with the VLT, and determined [Fe/H]$_{ZW}$ for \object{NGC\,1978} and 
\object{NGC\,1866}.  We adopted the estimated metallicities 
of \cite{hfs00} for 
\object{NGC\,1978} and \object{NGC\,1866} transformed to CG97 scale.

\cite{jt94} have observed spectra of stars at intermediate resolution in the 
clusters \object{NGC\,1850}, \object{NGC\,1854}, \object{NGC\,2004}, 
and \object{NGC\,2100}, comparing them with 
synthetic spectra in order to obtain metallicities, which we adopted. 
One star was observed in \object{NGC\,1850} and the error adopted 
arbitrarily corresponds to 20\%.

\cite{gcb95} have remeasured age-sensitive indices in the CMDs of 
\object{NGC\,1711}, \object{NGC\,1831}, \object{NGC\,1866}, 
\object{NGC\,1868}, \object{NGC\,2004}, \object{NGC\,2100}, 
\object{NGC\,2134}, \object{NGC\,2164} 
and  \object{NGC\,2214}, making it an homogeneous sample. 
We have checked consistency 
with our scale by comparing the age that they obtained for 
\object{NGC\,1866} with 
that by \cite{tfc99} \citep[also estimated by][]{gbd97}, finding both values 
similar ($0.15\pm0.06$\,Gyr and $0.15\pm0.05$\,Gyr, respectively). We have used
ages by \cite{gcb95} for the aforementioned clusters. 

\cite{drg00} determined ages and metallicities for \object{NGC\,1711}, 
\object{NGC\,2031} and \object{NGC\,2136} based on CCD Str\"omgren 
photometry. Different isochrone sets
were employed to derive ages. Judging by \object{NGC\,2031} and 
\object{NGC\,1711} 
($0.16\pm0.04$\,Gyr and $0.050\pm0.006$\,Gyr, respectively), also
in the \cite{gcb95} sample, the ages are on the same scale.
Their metallicities are in the ZW84 system, and therefore, the estimate for 
\object{NGC\,1711} has been transformed to the CG97 system.

For SMC clusters, the metallicities in the CG scale and homogeneous ages 
according to \cite{dh98} were adopted, except for \object{K\,28} \citep{psc01}
and for \object{NGC\,121} and \object{K\,3}, whose ages are from \cite{msf98}.

\section{The integrated spectra and equivalent widths}

EWs of metallic features (K\,Ca\,II, G\,band\,(CH)
and Mg\,I) and Balmer lines (H$\delta$, H$\gamma$ and  H$\beta$) were 
taken from  \cite{ba86b}, except for 
the LMC clusters \object{NGC\,1711}, \object{NGC\,1805}, \object{NGC\,1850}, 
\object{NGC\,1854}, \object{NGC\,1984} and
\object{NGC\,2011}, which were taken from \cite{sbc95}. This source provides
the EW of Mg\,I+MgH, which is a sum of three adjacent windows. We have 
measured in those spectra the central window (5156-5196\AA), which is the one
employed in the present work, according to
the definitions in \cite{ba86a}. EWs measured for \object{NGC\,6520} and 
\object{Mel\,105} were taken from \cite{sb93}. EWs measurements of all 
six windows
were carried out for \object{ESO\,93-SC08}, \object{NGC\,5606}, 
\object{NGC\,6253}, \object{NGC\,6540}, \object{K\,3} and \object{K\,28}.

We emphasize that the measurement of the EWs
follows a uniform procedure: first, the continuum placement
according to well-determined spectral fluxes is fitted and, second, the 
spectral windows as defined by \cite{ba86a} are fixed. Limits for the 
K\,Ca\,II, G band (CH), Mg\,I, H$\delta$, H$\gamma$ and  H$\beta$
spectral windows are, respectively, (3908-3952)\AA, (4284-4318)\AA, 
(5156-5196)\AA, (4082-4124)\AA, (4318-4364)\AA, and (4846-4884)\AA. Such a 
procedure has been applied consistently for all of the cluster sample, 
making the EWs from integrated spectra safely 
comparable and useful to study stellar populations in general. 
 
Before measuring the EWs, the spectra were set to the rest-frame according to 
the Doppler shift of H Balmer lines. Then, the spectra were normalized to 
F$_{\lambda}=1$ at 5870\AA~ and smoothed to the typical resolution of the 
database ($\approx10-15\AA$).

The EWs of H Balmer,  K\,Ca\,II, G\,band\,(CH) and 
Mg\,I\,(5167\,+\,5173\,+\,5184\AA) were measured using 
the IRAF task {\it splot}. Tables~\ref{tabmw_ew} and \ref{tabmc_ew}
present these measurements. Typical errors of 10\% on individual EW 
measurements were obtained by employing slightly different continua.

\begin{table}
\tiny
\caption[]{Equivalent widths: Milky Way clusters.}
\label{tabmw_ew}
\begin{tabular}{lrrrrrrr}
\hline
Window & K\,Ca\,II & G\,band\,CH &Mg\,I & H$\delta$ &H$\gamma$ &H$\beta$\\
 & & & & & & & \\
CLUSTER            & & & & & & \\
\hline
GLOBULAR\\
\hline
\object{NGC\,104 }      & 12.9 & 5.3 & 5.6 & 2.2 & 1.8 & 2.9\\
\object{NGC\,362 }      & 10.5 & 3.9 & 1.9 & 2.1 & 2.6 & 2.9\\
\object{NGC\,1851}      & 10.5 & 3.1 & 2.0 & 4.1 & 3.6 & 3.5\\
\object{NGC\,1904}      &  9.1 & 3.2 & 1.2 & 3.7 & 3.5 & 3.4\\
\object{NGC\,2808}      &  8.5 & 3.7 & 1.6 & 2.9 & 2.8 & 3.3\\
\object{NGC\,4590}      &  4.9 & 1.5 & 1.8 & 3.6 & 3.4 & 3.8\\
\object{NGC\,4833}      &  4.1 & 3.7 & 2.2 & 6.4 & 4.6 & 3.8\\
\object{NGC\,5024}      &  5.7 & 2.7 & 1.0 & 3.8 & 3.7 & 3.6\\
\object{NGC\,5824}      &  5.3 & 2.9 & 1.3 & 4.5 & 4.8 & 2.7\\
\object{NGC\,5927}      & 14.6 & 9.0 & 6.2 & 4.7 & 4.9 & 4.1\\
\object{NGC\,5946}      & 10.1 & 6.2 & 2.1 & 6.6 & 5.3 & 3.4\\
\object{NGC\,6093}      &  5.8 & 3.5 & 1.3 & 4.1 & 4.8 & 3.2\\
\object{NGC\,6139}      &  6.1 & 3.8 & 3.3 & 3.8 & 5.4 & 3.9\\
\object{NGC\,6171}      & 14.4 & 7.2 & 5.0 & 3.8 & 5.8 & 2.7\\
\object{NGC\,6287}      &  5.8 & 3.6 & 1.8 & 4.7 & 4.4 & 3.1\\
\object{NGC\,6293}      &  6.4 & 4.2 & 0.9 & 5.9 & 5.8 & 4.9\\
\object{NGC\,6304}      & 17.4 & 7.8 & 5.6 & 4.0 & 5.6 & 2.7\\
\object{NGC\,6316}      & 15.0 & 8.4 & 5.3 & 2.0 & 5.8 & 2.9\\
\object{NGC\,6356}      & 17.5 & 8.1 & 5.1 & 3.5 & 4.2 & 2.8\\
\object{NGC\,6388}      & 14.3 & 5.8 & 5.3 & 4.3 & 4.1 & 2.5\\
\object{NGC\,6401}      & 11.1 & 6.5 & 4.7 & 9.3 &10.8 & 4.2\\
\object{NGC\,6402}      &  5.4 & 5.9 & 3.2 & 6.0 & 6.7 & 3.9\\
\object{NGC\,6440}      & 17.1 & 9.3 & 7.8 & 7.2 & 7.4 & 4.8\\
\object{NGC\,6453}      &  6.6 & 4.9 & 1.7 & 4.3 & 4.9 & 3.9\\
\object{NGC\,6517}      &  9.8 & 6.5 & 0.3 & 7.4 & 6.6 & 3.5\\
\object{NGC\,6528}      & 15.9 & 8.3 & 8.4 & 5.2 & 6.3 & 4.3\\
\object{NGC\,6540}      & 14.8 & 7.5 & 5.3 & 2.7 & 3.7 & 3.2\\
\object{NGC\,6541}      &  6.7 & 3.1 & 1.4 & 3.8 & 2.6 & 4.0\\
\object{NGC\,6544}      &  8.4 & 4.9 & 3.3 & 3.4 &10.9 & 4.1\\
\object{NGC\,6553}      & 18.6 &14.0 & 8.5 & 0.4 & 5.4 & 6.2\\
\object{NGC\,6558}      & 10.5 & 4.8 & 3.0 & 6.3 & 4.1 & 4.9\\
\object{NGC\,6569}      & 13.7 & 7.7 & 5.5 & 5.2 & 6.5 & 2.3\\
\object{NGC\,6624}      & 15.5 & 6.3 & 5.4 & 3.6 & 4.2 & 2.5\\
\object{NGC\,6637}      & 14.0 & 8.0 & 3.2 & 2.2 & 3.6 & 1.8\\
\object{NGC\,6638}      & 12.0 & 5.6 & 3.0 & 3.7 & 4.8 & 4.2\\
\object{NGC\,6642}      &  8.5 & 7.7 & 4.4 & 5.4 & 5.7 & 4.3\\
\object{NGC\,6652}      & 11.4 & 6.4 & 3.5 & 2.9 & 3.6 & 2.7\\
\object{NGC\,6715}      &  8.8 & 4.9 & 2.8 & 3.6 & 4.4 & 3.0\\
\object{NGC\,6760}      & 15.1 & 7.6 & 6.6 & 8.8 & 4.4 & 3.0\\
\object{NGC\,6864}      & 11.2 & 4.2 & 3.2 & 3.2 & 3.3 & 3.8\\
\object{NGC\,7006}      & 10.4 & 5.0 & 2.2 & 5.0 & 4.8 & 3.0\\
\object{NGC\,7078}      &  5.4 & 2.4 & 0.9 & 3.2 & 4.0 & 2.5\\
\hline
\end{tabular}
\end{table}
\setcounter{table}{2}

\begin{table}
\tiny
\caption[]{Equivalent widths: Milky Way clusters (cont.)}
\begin{tabular}{lrrrrrrr}
\hline
Window & K\,Ca\,II & G\,band\,CH &Mg\,I & H$\delta$ &H$\gamma$ &H$\beta$\\
 & & & & & & & \\
CLUSTER            & & & & & & \\
\hline
OPEN\\
\hline
\object{NGC\,2158    }& 13.4 & 4.4 & 5.0 &10.1 & 8.3 & 7.4\\
\object{vdB-RN\,80   }&  0.6 & 0.3 & 1.2 & 6.5 & 5.8 & 5.3 \\
\object{NGC\,2368    }&  3.2 & 4.3 & 2.1 &10.7 &10.1 & 8.7  \\
\object{Berkeley\,75 }& 10.4 & 9.2 & 5.0 & 4.0 & 6.3 & 3.5\\
\object{Haffner\,7   }&  1.6 & 4.4 & 3.3 & 8.7 &15.4 & 7.3\\
\object{ESO\,429-SC13}&  1.3 & 3.4 & 2.6 & 8.7 &11.6 & 8.7\\
\object{NGC\,2660    }&  8.4 & 3.8 & 3.1 & 8.4 & 7.8 & 8.2\\
\object{UKS\,2       }&  7.7 & 5.4 & 3.3 & 9.3 & 6.9 & 5.7\\
\object{Ruprecht\,83 }&  2.3 & 1.6 & 1.4 &11.5 & 8.5 & 8.3\\
\object{Hogg\,3      }&  5.3 & 1.8 & 1.5 & 8.4 & 8.1 & 7.6\\
\object{NGC\,3293    }& -1.5 & 2.1 & 1.0 & 3.0 & 4.4 & 1.2 \\
\object{Bochum\,12   }&  4.5 & 2.6 & 2.3 & 7.6 & 8.6 & 6.4\\
\object{Pismis\,17   }&  2.0 & 2.5 & 0.5 & 5.0 & 6.9 & 1.4\\
\object{Hogg\,11     }&  2.7 &-0.2 & 0.3 & 4.5 & 3.5 & 3.6\\
\object{ESO\,93-SC08 }& 12.5 & 6.1 & 4.8 & 6.7 & 4.6 & 4.2\\
\object{MEL\,105     }&  2.3 & 2.4 & 1.3 &10.7 &11.0 & 9.6\\
\object{BH\,132      }&  4.6 & 1.1 & 2.9 &15.1 & 8.6 & 7.2 \\
\object{Hogg\,15     }&  0.0 & 0.3 & 0.7 & 5.7 & 3.9 & 3.8\\
\object{Pismis\,18   }&  6.4 & 4.8 & 2.8 & 7.5 & 6.3 & 6.4\\
\object{NGC\,5606    }&  1.5 & 0.4 & 0.5 & 5.1 & 4.3 & 3.6\\
\object{NGC\,5999    }&  3.3 & 3.0 & 3.2 &10.5 & 9.8 & 8.3\\
\object{NGC\,6031    }&  2.0 & 1.0 & 0.4 & 9.5 & 8.8 & 7.7\\
\object{Ruprecht\,119}&  0.9 & 1.1 & 1.3 & 6.9 & 5.0 & 3.8\\
\object{NGC\,6178    }&  0.5 & 0.2 & 0.2 & 7.6 & 5.5 & 5.3\\
\object{Lyng\aa\,11  }&  5.3 & 3.8 & 2.5 & 9.5 &10.1 & 7.6\\
\object{NGC\,6253    }& 11.8 & 4.9 & 7.4 & 8.7 & 3.0 & 6.4\\
\object{BH\,217      }&  2.3 & 2.3 & 0.6 & 6.4 & 5.5 & 3.3\\
\object{NGC\,6318    }&  3.7 & 4.6 & 1.0 & 8.8 & 9.8 & 4.8\\
\object{NGC\,6520    }&  3.0 & 3.5 & 2.3 & 7.4 & 8.7 & 6.6\\
\object{NGC\,6603    }&  5.6 & 3.4 & 2.1 &14.3 &14.3 &12.2\\
\object{Ruprecht\,144}&  1.3 & 2.0 & 0.9 & 9.4 &11.0 & 7.3\\
\object{NGC\,6705    }&  2.9 & 2.6 & 2.1 &10.5 &10.1 & 8.6\\
\object{NGC\,6756    }&  1.7 & 1.8 & 1.1 &11.2 & 8.6 & 8.7\\
\hline
\end{tabular}
\end{table}

\begin{table}
\tiny
\caption[]{Equivalent widths: Magellanic Cloud clusters.}
\label{tabmc_ew}
\begin{tabular}{lrrrrrrr}
\hline
Window & K\,Ca\,II & G\,band\,CH &Mg\,I & H$\delta$ &H$\gamma$ &H$\beta$\\
 & & & & & & & \\
CLUSTER            & & & & & & \\
\hline
LMC\\
\hline
\object{NGC\,1466    }&  2.6 & 3.1 & 3.0 & 7.0 & 6.5 & 4.5\\
\object{NGC\,1711    }&  1.4 & 1.1 & 1.2 & 7.0 & 6.8 & 5.5 \\
\object{NGC\,1783    }&  8.5 & 4.4 & 3.3 & 7.5 & 6.9 & 6.5\\
\object{NGC\,1805    }&  1.0 & 1.3 & 2.5 & 6.6 & 6.0 & 4.7\\
\object{NGC\,1831    }&  4.7 & 2.6 & 1.8 &14.6 &10.4 & 7.2\\
\object{NGC\,1850    }&  1.1 & 1.0 & 1.0 & 7.4 & 7.1 & 6.8\\
\object{NGC\,1854    }&  2.9 & 2.2 & 2.5 & 7.8 & 7.3 & 3.2\\
\object{NGC\,1856    }&  2.1 & 2.9 & 1.9 &11.6 &10.0 & 8.2\\
\object{NGC\,1866    }&  0.8 & 1.2 & 1.2 & 8.8 & 6.8 & 7.0\\
\object{NGC\,1868    }&  6.3 & 1.4 & 2.7 &10.9 & 7.7 & 7.9\\
\object{NGC\,1978    }& 11.1 & 4.5 & 3.1 & 3.7 & 4.8 & 5.4\\
\object{NGC\,1984    }&  0.9 & 1.5 & 1.0 & 4.2 & 5.3 & 2.2\\
\object{NGC\,2004    }&  1.2 & 0.0 & 2.0 & 4.3 & 3.2 & 3.5\\
\object{NGC\,2011    }&  0.8 & 0.9 & 1.4 & 2.6 & 3.9 & 2.0\\
\object{NGC\,2100    }&  2.5 & 0.4 & 1.5 & 4.1 & 3.5 & 1.8\\
\hline
SMC\\
\hline
\object{NGC\,121     }& 11.6 & 3.5 & 2.1 & 2.7 & 1.2 & 3.1\\
\object{NGC\,330     }&  0.5 & 0.5 & 0.0 & 5.5 & 4.1 & 4.6\\
\object{NGC\,419     }&  4.8 & 3.3 & 1.1 & 6.9 & 7.1 & 6.3\\
\object{K\,3         }&  6.8 & 6.2 & 3.4 & 5.4 & 3.9 & 4.5\\
\object{K\,28        }&  9.1 & 4.1 & 3.4 & 2.9 & 3.4 & 2.9\\
\hline
\end{tabular}
\end{table}

EWs of the selected three metallic features have been shown to be well 
correlated with metallicity for the old GGCs \citep{ba86a}. EWs of Balmer 
lines, being sensitive to the flux of turnoff stars for intermediate-age/young 
clusters, are expected to change with cluster age, reaching a maximum whenever
A-type stars dominate the turnoff \citep{ba86a}.

Taking into account these trends and, with the aim of finding spectral
indices with a higher sensibility to the cluster ages and metallicities, we 
analysed the behaviour of the sum of EWs of the three metallic lines and the 
three Balmer lines. As a by-product, the relative errors of these sums resulted
in $\approx7\%$ smaller errors than the individual spectral window. A similar 
approach has been shown to be useful in the discrimination of old and 
intermediate-age/young systems \citep{r82,dbc99}.

\section{Calibrating age and [Fe/H] using $\Sigma$EW(K+CH+Mg) and 
$\Sigma$EW(H${\delta}$+H${\gamma}$+H${\beta}$)}

\begin{figure}
 \resizebox{\hsize}{!}{\includegraphics{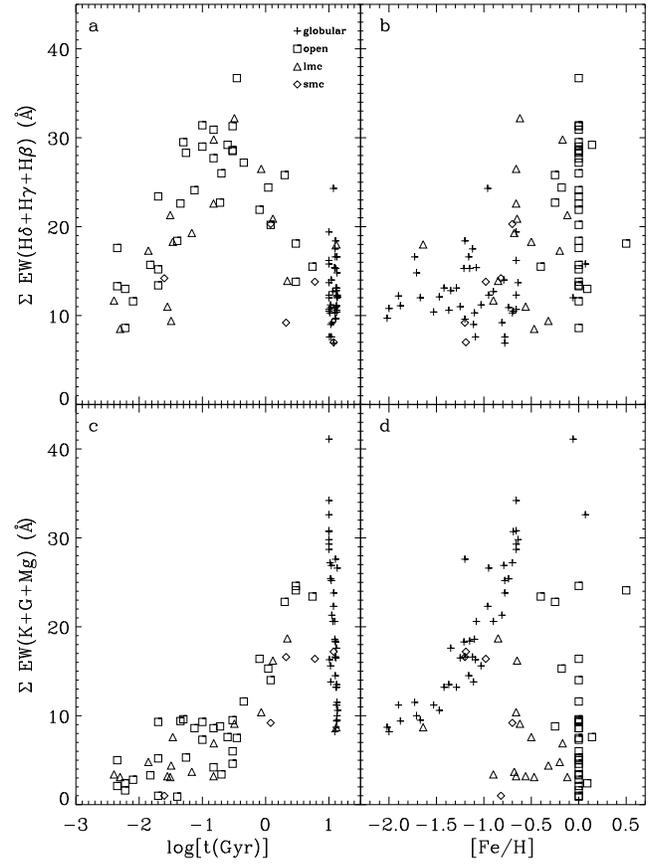}}
 \caption{Sum of metallic and Balmer H equivalent widths against 
 cluster parameters. Different symbols distinguish Milky Way and Magellanic 
 Clouds star clusters.}
 \label{ew_par}
\end{figure}

Fig.~\ref{ew_par} presents the EW sums against cluster age and 
metallicity. Different symbols represent clusters of different type or 
parent galaxy as indicated at the top of the upper-left panel. Error bars are 
not shown for clarity purposes. At a first glance, both metallic and Balmer 
line EW sums seem to be correlated with the age of clusters younger than 10 Gyr
(upper- and lowel-left panels). According to the lower right panel, the sum of 
metallic line EWs also appears to be sensitive to the GGC metallicities.
However, none of the EW sums is correlated with GGC ages - despite the fact
that they cover an appreciable age range ($t$ $\approx$ 10 - 14 Gyr)-, nor
with the metallicity of clusters younger than 10 Gyr.

\begin{figure}
 \resizebox{\hsize}{!}{\includegraphics{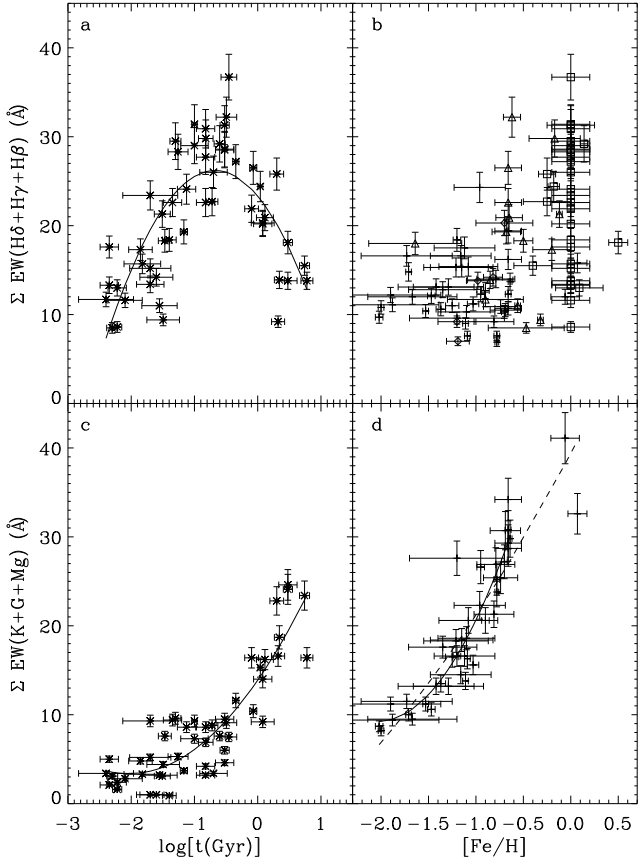}}
 \caption{Same as Fig.~\ref{ew_par}, including error bars, for selected 
 subsamples: {\bf(a)}~Balmer H equivalent width against age for clusters with 
 t $<10$\,Gyr; a 2nd order polynomial fitted to the data is superimposed; 
 {\bf(b)}~Balmer H EW against [Fe/H]; {\bf(c)}~Metallic features 
 EW against age for 
 clusters with t$<10$\,Gyr; a 2nd order polynomial fit is shown; 
 {\bf(d)}~Metallic 
 features EW against [Fe/H] for Galactic globular clusters; the continuous line
 corresponds to a 2nd order polynomial fitted to clusters with [Fe/H] $<-0.5$, 
 while the dashed line is a similar fit for the whole sample of GGCs.}
 \label{ew_par_fit}
\end{figure}

 According to the observed behaviours, we decided to fit both EW sums
in terms of $\log{\rm t}$ for clusters younger than 10 Gyr, and the metallic
line EW sum as a function of GGC metal content. Fig.~\ref{ew_par_fit} shows 
the same plots as Fig.~\ref{ew_par}, but including error bars and curve 
fittings superimposed on diagrams for which a correlation was found. Ranges of 
age and metallicity were selected as well to highlight their 
correlations with the EWs. For the relationship between
the $\Sigma{\rm EW(H\delta+H\gamma+H\beta)}$ and the cluster age (upper-left 
panel), we adjusted a quadratic polynomial given by the
expression: 

\begin{equation}
\label{ew_par_fita}
\Sigma{\rm EW(H\delta+H\gamma+H\beta)}=k_1+k_2.\log{\rm t(Gyr)}+k_3.(\log{\rm t(Gyr)})^2
\end{equation}

\noindent where $k_1$, $k_2$, and $k_3$ resulted to be 23.32$\pm$0.20, 
-8.56$\pm$0.35, and -6.35$\pm$0.18, respectively. For $\Sigma{\rm EW(K+G+Mg)}$
versus the cluster age, we fitted the equation:

\begin{equation}
\label{ew_par_fitb}
\Sigma{\rm EW(K+G+Mg)}=p_1+p_2.\log{\rm t(Gyr)}+p_3.(\log{\rm t(Gyr)})^2
\end{equation}

\noindent where $p_1$, $p_2$, and $p_3$ are 13.88$\pm$0.20, 
10.32$\pm$0.35, and 2.53$\pm$0.18, respectively. The root mean square errors
of eq.~\ref{ew_par_fita} and eq.~\ref{ew_par_fitb} turned out to be 4.8 and
2.9, respectively, the equations being valid in the range 
$-2.4<\log{\rm t(Gyr)}<0.8$ (see Fig.~\ref{ew_par_fit}a,c).

The curve fitting shown with solid line in Fig.~\ref{ew_par_fit}d,
i. e., the calibration including GGCs for which [Fe/H] $< -0.5$ is given
by the equation:

\begin{equation}
\label{ew_par_fitc}
\Sigma{\rm EW(K+G+Mg)}=q_1+q_2.{\rm [Fe/H]}+q_3.({\rm [Fe/H]})^2
\end{equation}

\noindent where $q_1=53.6\pm1.5$, $q_2=43.7\pm2.5$, $q_3=10.78\pm0.96$ 
and $\sigma$($\Sigma{\rm EW(K+G+Mg)}$)$=2.7$. The relationship is valid in 
the range $-2.0<{\rm [Fe/H]}<-0.65$. If the fit is performed using all 
GGCs in our sample ($-2.0<{\rm [Fe/H]}<0.07$), the coefficients for the 
expression:

\begin{equation}
\label{ew_par_fitd}
\Sigma{\rm EW(K+G+Mg)}=u_1+u_2.{\rm [Fe/H]}+u_3.({\rm [Fe/H]})^2
\end{equation}

\noindent give $u_1=39.40\pm0.63$, $u_2=20.1\pm1.1$,
$u_3=1.92\pm0.50$, and $\sigma$($\Sigma{\rm EW(K+G+Mg)}$)$=3.1$.

To facilitate the direct use of eqs. \ref{ew_par_fitb} 
to \ref{ew_par_fitd} we present below the corresponding 
inverted expressions. 

For clusters younger than $\log{\rm t(Gyr)}<0.8$:

\begin{equation}
\log{\rm t(Gyr)}=a_1+a_2.\Sigma{\rm EWm}+a_3.(\Sigma{\rm EWm})^2
\end{equation}

\noindent where $\Sigma{\rm EW(K+G+Mg)}$ was abbreviated to $\Sigma{\rm EWm}$ 
and the coefficients are $a_1=-2.18\pm0.38$,  $a_2=0.188\pm0.080$, 
$a_3=-0.0030\pm0.0032$, with $\sigma$($\log{\rm t(Gyr)}$)$=0.48$.

For GGCs with [Fe/H] $< -0.5$:

\begin{equation}
{\rm [Fe/H]}=b_1+b_2.\Sigma{\rm EWm}+b_3.(\Sigma{\rm EWm})^2
\end{equation}

\noindent with $b_1=-2.9\pm1.2$,  $b_2=0.14\pm0.14$, 
$b_3=-0.0023\pm0.0033$, and $\sigma$([Fe/H])$=0.14$.

And for all GGCs:

\begin{equation}
{\rm [Fe/H]}=c_1+c_2.\Sigma{\rm EWm}+c_3.(\Sigma{\rm EWm})^2
\end{equation}

\noindent with $c_1=-2.48\pm0.98$,  $c_2=0.088\pm0.097$, 
$c_3=-0.0008\pm0.0022$, and $\sigma$([Fe/H])$=0.16$.

\section{Diagnostic diagrams}

\begin{figure}
 \resizebox{\hsize}{!}{\includegraphics{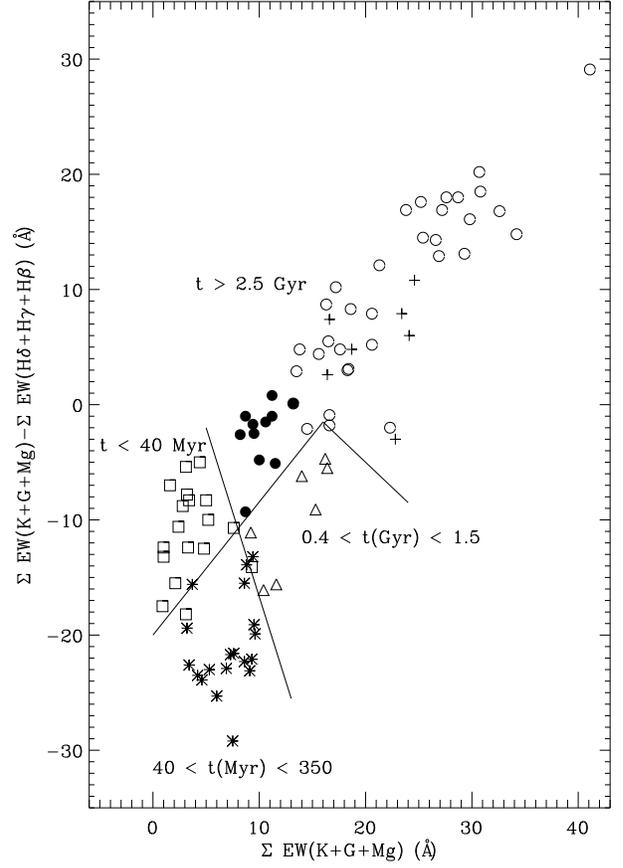}}
 \caption{The difference between EW of metallic features and EW of Balmer H 
 as a function of EW of metallic features. Different symbols and lines 
 discriminate clusters by age. Galactic globular clusters are indicated by
open circles ([Fe/H]$>-1.4$) and filled circles ([Fe/H]$\leq-1.4$) and 
intermediate-age clusters ($2.5 <$ t(Gyr) $<10$) are shown as plus signs.}
 \label{sm_shxsm}
\end{figure}

\begin{figure}
 \resizebox{\hsize}{!}{\includegraphics{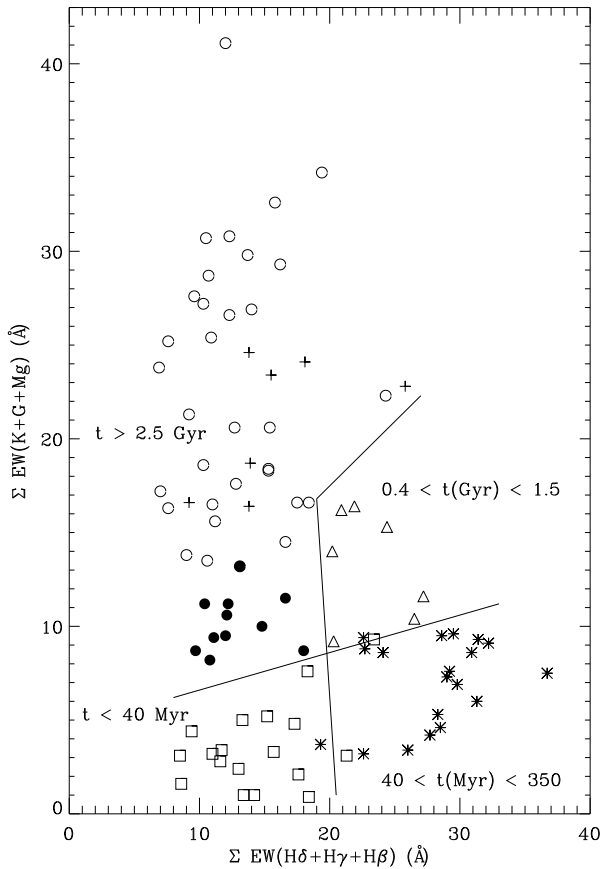}}
 \caption{Sum of metallic versus Balmer H EWs. Symbols and lines are as in 
 Fig.~\ref{sm_shxsm}.}
\label{smxsh}
\end{figure}

In order to provide a useful tool for stellar population studies,
we discuss in this section diagrams based on the two used EW sums.
After testing several combinations, we built two significant 
diagrams in terms of discriminating clusters of different ages.
Fig.~\ref{sm_shxsm} and Fig.~\ref{smxsh} show these diagrams, namely,
[$\Sigma{\rm EW(K+G+Mg)}-\Sigma{\rm EW(H\delta+H\gamma+H\beta)}$]~vs~$\Sigma{\rm EW(K+G+Mg)}$ and $\Sigma{\rm EW(K+G+Mg)}$~vs~$\Sigma{\rm EW(H\delta+H\gamma+H\beta)}$. 

Together with the calibrations presented in  Fig.~\ref{ew_par_fit},
these diagnostic diagrams may allow one to estimate ages for Galactic
and extragalactic star clusters, and overall metal abundances for
globular clusters, using their integrated spectra in the visible range. 
The behaviour of the EW sums and their combinations with cluster parameters
is very different above and below t$\approx$\,10\,Gyr.
$\Sigma{\rm EW(K+G+Mg)}$ is well determined by [Fe/H] for clusters with 
t $>$\,10\,Gyr, being independent of age within this range 
(see Fig.~\ref{ew_par_fit}d). For clusters with t $<$\,10\,Gyr, age is the 
dominant parameter on both $\Sigma{\rm EW(K+G+Mg)}$ (see 
Fig.~\ref{ew_par_fit}c) and $\Sigma{\rm EW(H\delta+H\gamma+H\beta)}$ (see 
Fig.~\ref{ew_par_fit}a).

On the basis that both EW sums are available and measured with an 
accuracy better than 10\%, we propose the following iterative procedure to 
obtain ages of star clusters: firstly, start using Figs.~\ref{sm_shxsm} and 
\ref{smxsh} to get a first estimate of the cluster age. Secondly, derive
calculated EW sums by using eqs.~\ref{ew_par_fita} 
and \ref{ew_par_fitb} with the estimated average age. Thirdly, use these
calculated EWs values to derive an improved age estimate from 
Figs.~\ref{sm_shxsm} and \ref{smxsh}. The procedure can be iterated until
a fixed difference between the updated and previous EWs values is 
reached. 
Usually, no more than one iteration is needed to reach a precision of 
0.1 \AA~ 
in $\Sigma{\rm EW(H\delta+H\gamma+H\beta)}$ and $\Sigma{\rm EW(K+G+Mg)}$. 
Note that globular age-like clusters fall outside the calibrating age range. 
However, globular age-like clusters can be recognized in the diagnostic 
diagrams.  Metal-poor GGCs with [Fe/H]$\leq-1.4$, represented by 
filled circles in Figs.~\ref{sm_shxsm} and \ref{smxsh}, lie in a distinct
region to that occupied by younger clusters. On the other hand, 
metal-richer globular clusters, represented by open circles in 
Figs.~\ref{sm_shxsm} and \ref{smxsh},
are approximately distributed in both figures over the same regions as 
intermediate-age clusters ($2.5<$ t(Gyr) $<10$). Consequently, one could be 
dealing with a metal-rich globular or an intermediate-age cluster without 
noticing it. For confirmed globular age-like clusters, 
metallicities can be obtained from the calibrations of Fig.~\ref{ew_par_fit}d.

\section{Concluding remarks}

We present new calibrations and diagnostic diagrams based on visible
integrated spectral features, which will allow one to derive ages for Galactic
and extragalactic star clusters, and also overall metallicities for clusters 
older than 10 Gyr. For that purpose, we first searched the literature for 
star clusters with integrated spectra and reliable determinations of their ages
and metallicities. This task led us to compile the largest sample of star
clusters, covering wide ranges of age and metallicity and with
the required integrated spectra available. We carried out an analysis as 
carefully as possible of ages and metal abundances of each selected 
object, in order to put them into homogeneous scales. As far as we are 
aware, this is the 
first time that such a sample of star clusters with homogeneous ages and 
metallicities is provided.

Using these homogeneous scales, we investigated the behaviours of different
integrated spectral indices as a function of cluster age and metallicity. The 
spectral indices were built on the basis of the measurements 
of EWs of K\,Ca\,II, 
G\,band\,(CH) and Mg\,I metallic lines, and H$\delta$, H$\gamma$, and H$\beta$ 
Balmer lines. The fact that all the EWs have been measured following the same 
precepts also guarantees homogeneity within the data. From the analysis, we 
found that both sums of the metallic and Balmer H line EWs, 
separately, are good 
indicators of cluster ages for objects younger than 10 Gyr. Likewise, 
$\Sigma{\rm EW(K+G+Mg)}$ is useful for metallicity determinations of star 
clusters older than 10 Gyr. The sensitivity of the suggested 
integrated spectral indices to cluster age or metallicity in the respective
age domain does not appear to be degenerated by the counterpart
parameter, in the sense that the EW sum, which is an age indicator, is not
a metallicity index in the corresponding age range, and vice versa.

Finally, we propose an iterative procedure for estimating star cluster ages
from two new diagnostic diagrams and two calibrations in terms of the 
mentioned EW sums. The method allows one to estimate cluster ages with an 
internal precision better than 10\%. For globular age-like clusters, we 
obtained a calibration of $\Sigma{\rm EW(K+G+Mg)}$ as a function
of the iron-to-hydrogen ratio. We foresee that this 
technique will serve to
estimate ages and metallicities of relatively faint clusters now reacheable 
with the new generation of telescopes and instruments.

\begin{acknowledgements}

We thank the referee, Dr. Raffaele G. Gratton, for helping to
improve this paper.
This work was partially supported by the Argentinian
institutions CONICET, and Agencia Nacional de Promoci\'on Cient\'{\i}fica y
Tecnol\'ogica (ANPCyT). 

\end{acknowledgements}

\bibliographystyle{apj}


\end{document}